\documentclass{PoS}

\usepackage{amsmath}
\usepackage{psfrag}

\title{Latest lattice results of $\mathcal{N}=1$ supersymmetric Yang-Mills theory with some 
topological insights}

\ShortTitle{Latest lattice results of $\mathcal{N}=1$ supersymmetric Yang-Mills theory}

\author{\speaker{Pietro Giudice}, Gernot M\"unster, Umut D. \"Ozugurel, 
  Stefano Piemonte,  
  
  Dirk Sandbrink\\
  Universit\"at M\"unster, Institut f\"ur Theoretische Physik, \\
  Wilhelm-Klemm-Str. 9, D-48149 M\"unster, Germany\\
  E-mail: \email{p.giudice@uni-muenster.de, munsteg@uni-muenster.de, 
oezugurel@uni-muenster.de, spiemonte@uni-muenster.de, 
dirk.sandbrink@uni-muenster.de}}

\author{Georg Bergner\\
  Universit\"at Frankfurt, Institut f\"ur Theoretische Physik, \\
  Max-von-Laue-Str. 1, D-60438 Frankfurt am Main, Germany\\
  E-mail: \email{bergner@th.physik.uni-frankfurt.de}
}

\author{Istvan Montvay\\
  Deutsches Elektronen-Synchrotron DESY, \\
  Notkestr. 85, D-22603 Hamburg, Germany\\
  E-mail: \email{montvay@mail.desy.de}
}

\abstract{
We summarise the latest results of our collaboration concerning $\mathcal{N}=1$
supersymmetric 
Yang-Mills theory in four dimensions on the lattice. We investigate the expected 
formation of supersymmetric multiplets of the lightest particles and the behaviour 
of the topological susceptibility approaching the supersymmetric limit of the theory.
}

\FullConference{The 32nd International Symposium on Lattice Field Theory\\
  23-28 June, 2014\\
  Columbia University New York, NY}

\newcommand{\aetap}{\text{a--}\eta'}
\newcommand{\api}{\text{a--}\pi}
\newcommand{\afn}{\text{a--}f_0}
\newcommand{\sapi}{\text{a}\mbox{-}\pi}

\newcommand{\tr}[1]{\ensuremath{\mathrm{Tr}} \left[{#1}\right]}

\newcommand{\beq}{\begin{equation}}
\newcommand{\eeq}{\end{equation}}
\newcommand{\bea}{\begin{eqnarray}}
\newcommand{\eea}{\end{eqnarray}}

\begin{document}

\section{Introduction}
Supersymmetry (SUSY) has proved to be a powerful concept which has been explored 
by physicists in different contexts for decades. 
It can be seen as an extension of the Poincar\'e symmetry of space-time, realised by
the introduction of supercharges, {\it i.e.} the generators of supersymmetry transformations.
Supercharges are operators which transform bosons into fermions, and vice versa.

In high energy physics SUSY is well known as a promising extension 
of the Standard Model, with strong support at both mathematical and physical level. 
SUSY can be realised in many different ways. The Large Hadron Collider (LHC) 
can probe extensively the low-energy realisation of some of them.
The first run of the LHC has restricted significantly
the parameter space of various SUSY models. Moreover, the non-observation of SUSY
states has shifted the SUSY partner masses into the TeV region.
As a consequence, even what was the first motivation of its introduction, {\it i.e.}
the possibility of solving the hierarchy problem, has been questioned.
Still, the last word about the realisation of this symmetry in nature, at least 
in this context, has not been said~\cite{Antoniadis:2014eea}. 

Moreover, SUSY is the key ingredient in many other areas of research. 
When \emph{local} supersymmetry is imposed, a new field theory is obtained where 
supersymmetry and general relativity live together in what we call supergravity.
SUSY has been incorporated in string theory, extending the previous bosonic string theory,
including fermionic degrees of freedom and originating the so called 
superstring theory.
In physical cosmology it is used to explain the presence of a small but nonzero 
cosmological constant.
It has been added in quantum mechanics before as an attempt to study the  
consequences of SUSY in a simpler setting, but later as an interesting topic by 
itself~\cite{Cooper:2001zd}.
There are applications in condensed matter physics in studying disordered   
and mesoscopic systems~\cite{Efetov:1997fw}. It is also used in  optical physics to tackle 
various theoretical problems~\cite{El-Ganainy:2013sda,Miri:2013kha}.
It is clear then that the study of the properties of supersymmetric theories, in 
particular the non-perturbative ones, continues to be of extreme interest.

The simplest non-abelian supersymmetric gauge theory, which is studied in this work, 
is the $\mathcal{N}=1$ 
supersymmetric Yang-Mills (SYM) theory with gauge group $SU(2)$. 
It describes the interaction between gluons and gluinos. 
The Lagrangian looks like the one of QCD with only one flavour, except that
in this theory the fermion field transforms in the adjoint representation
and it is a Majorana field. 
Usually a mass term is considered, which breaks SUSY softly. When the mass term is 
zero supersymmetry is predicted to be unbroken, even in the quantised theory~\cite{Witten:1982df}.

Like in QCD, the theory in SYM is asymptotically free at high energies and 
becomes stronly coupled in the infrared limit. Due to confinement, the spectrum of particles is 
expected to consist of colourless bound states. If supersymmetry is
unbroken the particles should belong to mass degenerate SUSY multiplets.

Many predictions concerning the properties of SYM theories are based on 
perturbation theory or semiclassical methods. However, some important properties 
are of a non-perturbative nature. The first predictions
on the spectrum of the theory were possible exploiting the fact that the symmetries 
of the theory constrain the form of the low-energy effective 
actions~\cite{Veneziano:1982ah, Farrar:1997fn}.
Verifying the formation of the predicted supermultiplets is a central task of our 
investigations.

Some important results have already been obtained by our collaboration in previous 
studies in the framework of a lattice-regularised version of SYM, 
see Refs.~\cite{Bergner:2013nwa,Bergner:2012rv,Demmouche:2010sf}. 
We have found that a rather small lattice spacing is necessary to investigate
the restoration of SUSY. 
In this work we have added the results of a further, even smaller,
lattice spacing. Due to the small lattice needed to reduce the supersymmetry breaking,
a closer look at the topological properties is required. 
In particular, some results regarding the topological susceptibility are presented.

\section{Chiral symmetry, SUSY, and continuum limit}
\label{sec:chiralcont}

As discussed in Ref.~\cite{Curci:1986sm, Suzuki:2012pc}, SUSY gauge theories can be studied on the lattice.
The main idea is that, rather than trying to have some version of SUSY on the 
lattice, which can be realized only in a non-local way,
one should only require that it is recovered in the continuum limit.
The conclusion of the two papers is that, in the continuum limit, the chiral limit defines the SUSY 
point and vice versa.

A fine tuning of the gluino mass $m_g$ is sufficient to approach supersymmetry in the continuum theory. 
This tuning is efficiently done by means of the mass of an \emph{unphysical} particle: the 
adjoint pion $\sapi$. Practically, the adjoint pion is defined by the connected contribution of the
correlator of the $\aetap$ particle. 
It has been suggested~\cite{Veneziano:1982ah} that, in the OZI approximation, 
the adjoint pion mass should vanish for a massless gluino. This has been then 
proved in a more formal way, using a partially quenched setup~\cite{Munster:2014cja},
arriving at the important conclusion that $m_{\api}^2~\propto~m_g$.

The strategy we follow to reach these limits consists of two steps: 
in the first, we fix the lattice spacing, {\it i.e.}
we run our simulations at fixed $\beta$ (the inverse of the coupling constant) 
and using several (3 or 4) values of the mass parameter $\kappa$, we extrapolate our results to the
chiral limit. In the second step, we extrapolate to the continuum limit, repeating
the first step for 3 or 4 values of $\beta$.

\section{Fixing the scale: $r_0$ and $w_0$}
\label{sec:scale}

The results the collaboration presented in Ref.~\cite{Demmouche:2010sf} were 
characterised by the constant $\beta=1.60$. A rather large gap, between fermionic and bosonic 
masses inside the same supermultiplet, was obsterved.
In the following we decreased the value of the lattice spacing by $\sim 40 \%$, increasing
the value of $\beta$ to 1.75. We presented the results 
in Refs.~\cite{Bergner:2013nwa, Bergner:2013jia} and for the first time we had 
some indications of a restoration of SUSY in the theory we are studying.
The results we present in this paper have been obtained on a $32^3\times64$ lattice and
they are characterised by $\beta=1.90$, which 
means a further reduction of the value of the lattice spacing 
by $\sim 30 \%$. Comparing the scale of this theory with that of QCD, we determined 
the value of the lattice spacing to be $a=0.03610(65)$~fm.
We are now very close to the continuum limit and the fundamental
picture starts to emerge.
\begin{figure}[hbt]
\begin{center}
\includegraphics[width=0.45\hsize,angle=0]{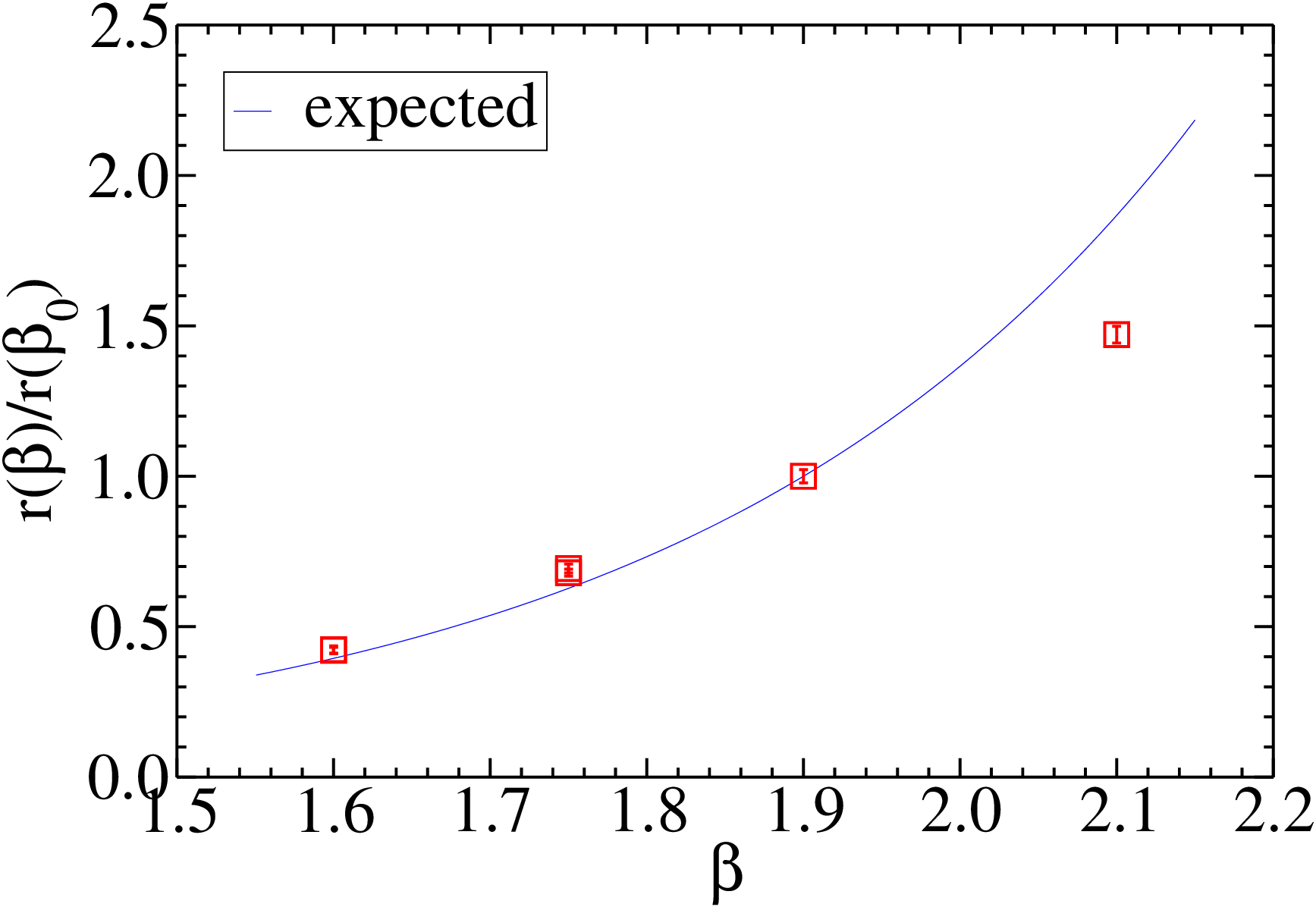}
\vspace{-6mm}
\end{center}
 \caption{Comparison of the Sommer parameter data, normalised to $\beta=1.90$, with the
expected value determined from the analytical $\beta$-function.}
 \label{fig:NSVZ}
\end{figure}
Because results with different lattice spacings have been obtained
it is now crucial to determine very accurately a scale so that all our
results can be compared and extrapolations to the continuum limit can be
carried out.
So far the scale has been determined using the Sommer parameter $r_0$~\cite{Sommer:1993ce}.
It is determined from the static quark potential, improving the signal using
APE smearing. The method requires two consecutive fitting procedures. 
Overall the method can be characterised by a few systematic errors which
sometimes are not easily taken into account. 

To see how good our determinations are, we compared the values of $r_0$
with the expected scaling. 
The $\beta$-function for SYM has been determined 
analytically~\cite{Novikov:1983uc}:
\begin{equation}
\beta(g)=-\frac{g^3}{16 \pi^2} \frac{3 N_c}{1-\frac{g^2 N_c}{8 \pi^2} } \ ,
\label{eq:NSVZ}
\end{equation}
where $N_c$ is the number of colours and $g$ the coupling constant.
The first two terms in the coupling constant expansion of the 
$\beta$-function are universal, namely scheme independent.

In Fig.~\ref{fig:NSVZ} we plot the ratio $a(\beta_0)/a(\beta)$, 
with $\beta_0=1.90$,
determined using a second order expansion in $g$ to calculate the integral of the $\beta$-function,
together with our numerical determination of the ratio of the Sommer parameters:  
$r(\beta)/r(\beta_0) \equiv a(\beta_0)/a(\beta)$. The result is pretty good: the first
three points, taking in account the errors, are in reasonable agreement with the theoretical expectation. 
The fourth point, which is our preliminary result for $\beta=2.10$,
has still some strong systematic error which will be discussed later in Sec.~\ref{sec:topology}.

An important improvement in the analysis of our data has been the
determination of the parameter $w_0$~\cite{Borsanyi:2012zs}, determined 
by Wilson flow~\cite{Luscher:2010iy}, to fix the scale. 
This parameter does not suffer from the systematic uncertainties
which are present in the Sommer parameter, providing a more reliable
comparison of our results. A plot similar to Fig.~\ref{fig:NSVZ} has been obtained
confirming the previous comments. 
More details will be presented in an upcoming paper.

\section{Light particle spectrum}
\label{sec:spectrum}

The low-lying spectrum of particles has been predicted by means of effective Lagrangians.
It consists of colour neutral bound states of gluons and gluinos, forming supermultiplets:
glueballs $gg$, gluinoballs (mesons) $\tilde{g}\tilde{g}$ and gluino-glueballs $\tilde{g}g$.

In Ref.~\cite{Veneziano:1982ah} interpolating operators for pure gluonic states have not been 
included, and only one supermultiplet was described.
It consists of a scalar ($0^+$ gluinoball: $\afn \sim \bar\lambda \lambda$), 
a pseudoscalar ($0^-$ gluinoball: $\aetap \sim \bar\lambda \gamma_5 \lambda$),
and a Majorana fermion (spin $1/2$ gluino-glueball: 
$\chi \sim \sigma^{\mu \nu} \tr{ F_{\mu \nu} \lambda }  $).

The effective Lagrangian of Ref.~\cite{Veneziano:1982ah} was generalised in Ref.~\cite{Farrar:1997fn}.
In addition to the first chiral supermultiplet a new one
appears:  a $0^-$ glueball, a $0^+$ glueball, and again a gluino-glueball.

As stressed by the authors, neither of these supermultiplets contain 
\emph{pure} gluino-gluino, gluino-gluon or gluon-gluon bound states. 
As a matter of fact, the physical excitations are mixed states of them: actually
in the limit when there is no mixing the two supermultiplets are degenerate.
This fact can have important consequences for the interpretation of the numerical results:
{\it e.g.} analysing a pure gluonic operator does not imply that we are determining the spectrum
of a gluon-gluon bound state.
In Fig.~\ref{fig:spectrum} four bound states are plotted, and their chiral limit, linearly 
extrapolated, is shown.
The mass of the gluino-glue and the $\aetap$ are the ones extrapolated with a better 
precision (relative error $\sim 10 \%$ and $\sim 15 \%$ respectively), 
than there is the $\afn$ and the scalar glueball (both with a relative error of $\sim 30 \%$).
\FIGURE{
  \psfrag{gtg}{\footnotesize{\textbf{$\tilde{g}g$}}}
  \psfrag{gg0p00}{\footnotesize{\textbf{$gg \ 0^+$}}}
  \includegraphics[width=6.1cm]{./figures/spectrum_01082014_2.eps}
  \caption{Spectrum of the theory at $\beta=1.90$ for four values of the $\api$ mass 
    and in the extrapolated chiral limit.}
  \label{fig:spectrum}
}
From this figure it is now clear that the three bound states belonging to
the first supermultiplet are degenerate within the error bars. 
Actually also the mass we determine studying the pure $0^{+}$ glueball
operator is compatible with the other three masses. 

In contrast, from our preliminary results the mass 
of the pseudoscalar glueball $0^-$ is almost two times the mass of the first supermultiplet.
A possible explanation could be that due to mixing the $0^{+}$ glueball operator 
has a significant overlap with the scalar state of the lower supermultiplet.
As a consequence, studying the corresponding correlator
at large euclidean-time distance, we find again the mass of the low-lying supermultiplet
particle with the same quantum number, {\it i.e.} the $\afn$.

Note that, contrary to what was assumed in Ref.~\cite{Farrar:1997fn}, 
this would imply that the lighter supermultiplet is gluinoball-like
and the higher is glueball-like.


\section{Topological susceptibility}
\label{sec:topology}

$\mathcal{N}=1$  SYM is characterised by the presence of topological sectors.
One of the greatest problem with this kind of theories, when local update algorithms
are used as in our case, is that the simulation may get stuck inside the same
topological sector. The transition between different topological sectors is suppressed
going closer to the continuum limit. It is therefore necessary to verify the shape
and the position of the distribution of the topological charge in every set of configurations generated.

In the case of QCD the topological susceptibility is a commonly studied observable, a quantity
which reflects the dependence of the vacuum energy on the vacuum angle. 
This quantity has not yet been studied intensively in SUSY models.
An exception is Ref.~\cite{Gorsky:2002wt}, where the topological susceptibility is 
discussed in the context of the orbifold equivalence.
The relevant result for us is that
the topological susceptibility is expected to go to zero proportionally to the quark mass, {\it i.e.}
proportionally to the square of the adjoint pion mass, according to the discussion 
in Sec.~\ref{sec:chiralcont}. This behaviour has been verified in our analysis.
\begin{figure}[hbt]
  \begin{center}
    \includegraphics[width=0.47\hsize,angle=0]{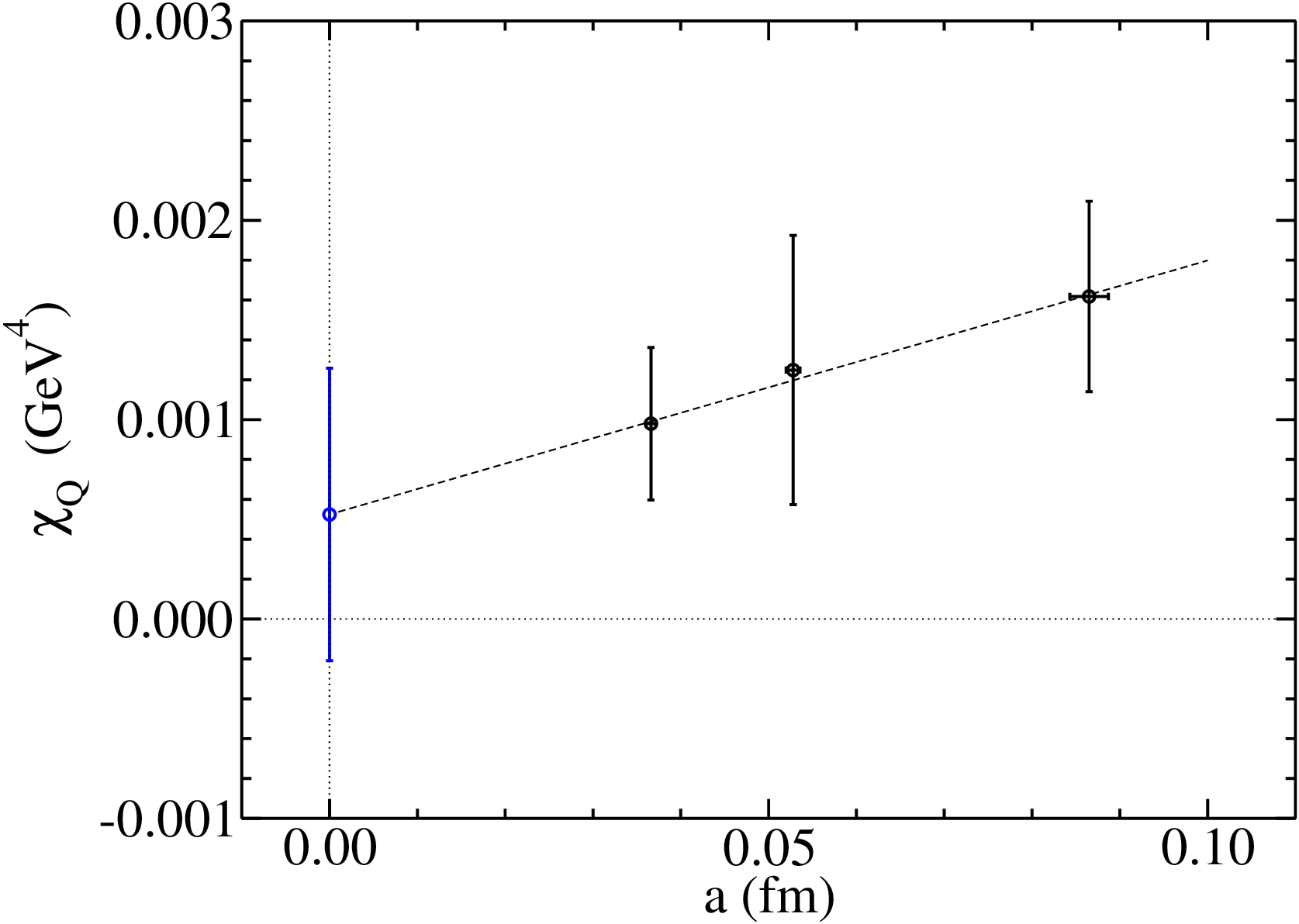}
    \vspace{-6mm}
  \end{center}
  \caption{Topological susceptibility extrapolated to the continuum limit. 
    Each point represents the chiral extrapolated value obtained fitting at least 3 values of $\kappa$.
    The conversion in dimensionful unit has been done using the value of the 
    Sommer parameter used in QCD: $r_0=0.5$~fm.}
  \label{fig:topolsusc}
\end{figure}

The topological charge $Q$, discussed in this paper, is defined as follows~\cite{Bonati:2014tqa}:
\beq
Q=\mbox{round}(\alpha Q_L) \ ,
\eeq
where $\mbox{round}(x)$ denotes the closest integer to $x$ and $Q_L$ is the
lattice definition in terms of smeared plaquettes. 
The scaling factor $\alpha$, which has been proved to improve the quality of the charge 
distributions, is determined minimising the quantity:
\beq
\langle \left( \alpha Q_L - \mbox{round}(\alpha Q_L) \right)^2 \rangle \ .
\eeq
The susceptibility of the topological charge is then defined by:
\beq
\chi_Q=\frac{1}{V} \left( \langle Q^2 \rangle - \langle Q \rangle^2  \right) \ .
\eeq
The lattice definition of the topological charge is affected by 
UV fluctuations, and as a consequence it does not take integer values.
This means that it is necessary to introduce a multiplicative renormalisation
and even an additive one for the topological susceptibility.
The way we deal with such renormalisations is based on smoothing methods: initially we compared
APE, HYP and stout smearing, and for the final measurements we focussed on APE smearing.

In Fig.~\ref{fig:topolsusc} the topological susceptibility is plotted against 
the lattice spacing as determined for our three values of $\beta$, and
its value extrapolated linearly to the continuum limit. 
A non-vanishing slope is not unexpected, because the fermion part of the action 
is improved only by applying a few levels of stout smearing to the link variables, 
and even using the tree-level Symanzik improved gauge action a linear dependence of 
observables on the lattice spacing is not prevented.
It should be noted that the extrapolated value is compatible with zero as expected.

The fact that in Fig.~\ref{fig:NSVZ} the point at $\beta=2.1$ is far below the expected 
value is related to the freezing of topology at that value of the lattice spacing, 
$a\sim0.019$~fm in QCD units, namely the scale parameter is dependent on the topological 
charge. 
We expect, however, that this difficulty can be overcome by optimising the
parameters ({\it e.g.} trajectory length) of our updating algorithm \cite{MontvayScholz}.

\section{Conclusions and outlooks}

We have presented our latest results on $\mathcal{N}=1$ supersymmetric Yang-Mills 
theory with gauge group $SU(2)$.
We are now able to see the degeneracy of the first supermultiplet, 
in accordance with the existence of a SUSY limit of the theory.

Some issues related to the mixing of 
the states, and therefore with the content of the second supermultiplet, have to be clarified.
We have started a systematic analysis of some topological properties of 
the theory, shedding light on this less known aspect of SUSY models.
Recently we have also started to explore the finite temperature 
properties of this theory~\cite{Bergner:2014saa}.

\section*{Acknowledgements}

This project is supported by the German Science Foundation (DFG) under
contract Mu 757/16.
The authors gratefully acknowledge the computing time granted by the John
von Neumann Institute for Computing (NIC) and provided on the supercomputers
JUQUEEN and
JUROPA at J\"ulich Supercomputing Centre (JSC).
Further computing time has been   
provided by the compute cluster PALMA of the University of M\"unster. 


\end{document}